\documentclass[aps,twocolumn,showpacs]{revtex4}
\usepackage{graphicx}

\begin{document}

\title{Universal Second-Order Phase Transition from Integrability to Chaos}

\author{$^1$Edson D.\ Leonel, $^1$Mayla A. M. de Almeida, $^2$Juan Pedro Tarigo, $^2$Arturo C. Marti, $^3$Diego F. M. Oliveira}

\affiliation{$^1$Departamento de F\'isica, Unesp - Universidade Estadual Paulista - 
Av.24A. 1515, 13506-700, Rio Claro, SP, Brazil\\
$^2$Facultad de Ciencias, Universidad de la Rep\'ublica, Igua 4225, Montevideo, Uruguay\\
$^3$School of Electrical Engineering and Computer Science, University of North Dakota, Grand Forks, Avenue Stop 8357, 58202, ND, USA}

\date{\today} \widetext


\begin{abstract}
We report a dynamical phase transition from integrability to non-integrability in a simple oval-like billiard with boundary $R(\theta)=1+\epsilon\cos(p\theta)$. For $\epsilon=0$, the phase space is {\it foliated} by invariant curves corresponding to periodic or quasiperiodic motion, whereas for small $\epsilon$ a thin chaotic layer separates rotational and librational trajectories. As $\epsilon$ increases, this layer grows according to a well-defined scaling law whose chaotic dispersion follows $\omega_{\rm rms,sat}\sim\epsilon^{\tilde{\alpha}}$, where the exponent $\tilde{\alpha}$ coincides with those of the Fermi-Ulam model, periodically corrugated waveguides, and a family of discrete mappings, revealing a universal mechanism for the onset of chaos in weakly perturbed integrable systems. The deviation of the reflection angle in the billiard, $\omega_{\rm rms,sat}$, acts as an order parameter: it vanishes continuously as $\epsilon\to 0$, signalling an ordered (integrable) phase, while its susceptibility $\chi=d\omega_{\rm rms,sat}/d\epsilon$ diverges, indicating a second-order phase transition. A symmetry breaking and an analytically solvable diffusion process complete the near-critical phenomenology. These results establish a unified framework for the emergence of chaos from integrability.
\end{abstract}

\maketitle

Chaos in classical billiards has long served as a paradigm in Hamiltonian dynamics, statistical mechanics, and nonlinear science \cite{r1,r2,r3}. Although the global structures of integrable \cite{r1}, ergodic \cite{r2,add5}, and mixed billiards \cite{diego,add6,add7} are well characterised, a fundamental question remains open: how does chaos emerge from integrability as a control parameter is varied, and can this process display the hallmarks of a phase transition, including order parameters, susceptibilities, and universality? If so, is any symmetry breaking involved? What is the elementary excitation that drives the dynamics to diffuse chaotically? Are there topological defects in phase space that influence the diffusion of particles?

Here we address these questions in an oval-like billiard with boundary $R(\theta)=1+\epsilon\cos(p\theta)$. For $\epsilon=0$ the system is integrable and the phase space is foliated \cite{r1}, meaning that the reflection angle remains constant, leading to periodic dynamics when the angle is commensurable with $\pi$ or quasiperiodic dynamics otherwise. For small $\epsilon$, a thin chaotic layer appears separating rotational from librational motion. As $\epsilon$ increases, this layer grows in width and eventually occupies almost the entire chaotic region, thereby revealing how chaos progressively develops in phase space as the control parameter increases. We show that the width of this layer obeys a scaling law $\sim\epsilon^{\tilde{\alpha}}$, where the critical exponent $\tilde{\alpha}$ coincides with those of the Fermi--Ulam model \cite{leonel1}, periodically corrugated waveguides \cite{leonel2}, and a family of two-dimensional, nonlinear and area-preserving mappings \cite{leonel3} whose angle diverges in the limit of vanishing action. This demonstrates the existence of a nontrivial universality class governing the onset of chaos in weakly perturbed integrable systems.

To characterise this transition, we analyse the deviation of the reflection angle around its mean value, $\omega_{\rm rms,sat}$, which acts as an order parameter \cite{r4}. As $\epsilon\to 0$, $\omega_{\rm rms,sat}$ vanishes continuously, while its susceptibility $\chi=d\omega_{\rm rms,sat}/d\epsilon$ diverges. This behavior \cite{r5,r6,r7} mirrors that of a second-order phase transition: a continuous suppression of chaotic fluctuations accompanied by a divergent response function. Stickiness \cite{r8}, island remnants \cite{r9}, and symmetry breaking \cite{r10} emerge near criticality, while the diffusion inside the chaotic layer is accurately captured by an analytically solvable diffusion equation \cite{r11}. These results establish the integrability--chaos transition in classical billiards as a genuine second-order phase transition with universal features, providing a unified framework for chaotic (stochastic) layers \cite{r12,r13}, anomalous transport \cite{r14}, and critical behavior \cite{r15} near the onset of chaos.

\begin{figure}[t]
\centerline{(a)\includegraphics[width=0.47\linewidth]{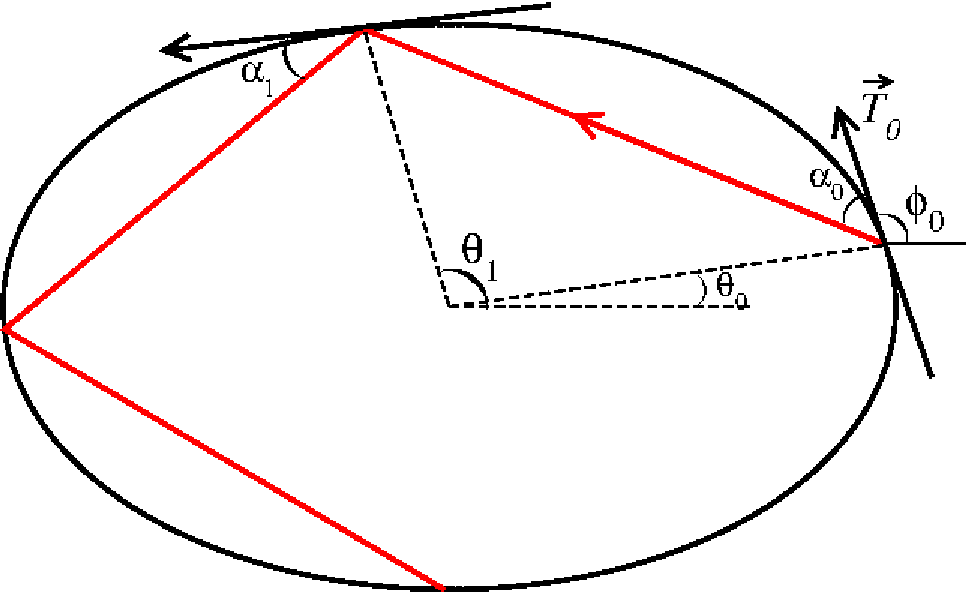}
(b)\includegraphics[width=0.47\linewidth]{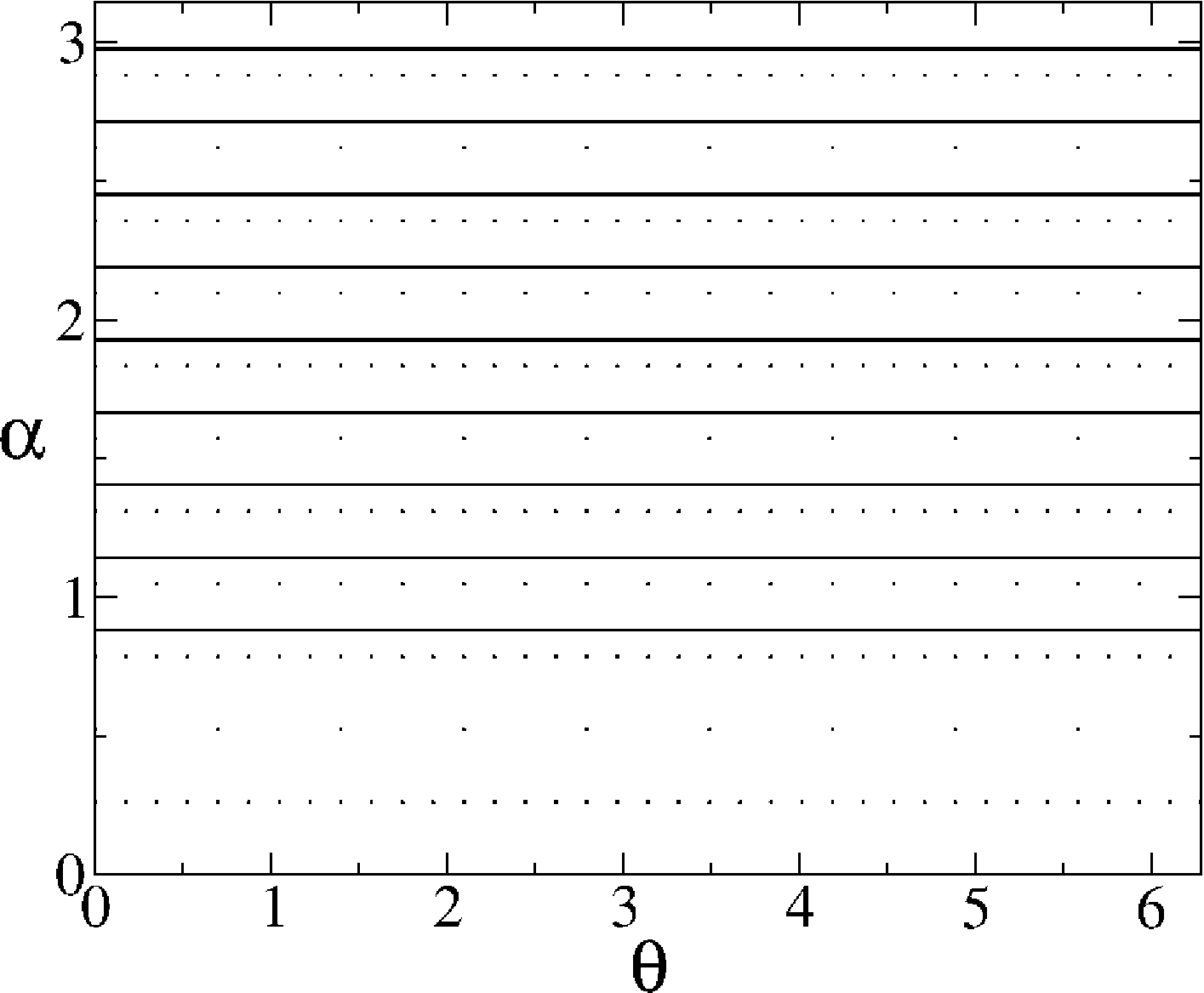}}\vspace{0.5cm}
\centerline{(c)\includegraphics[width=0.47\linewidth]{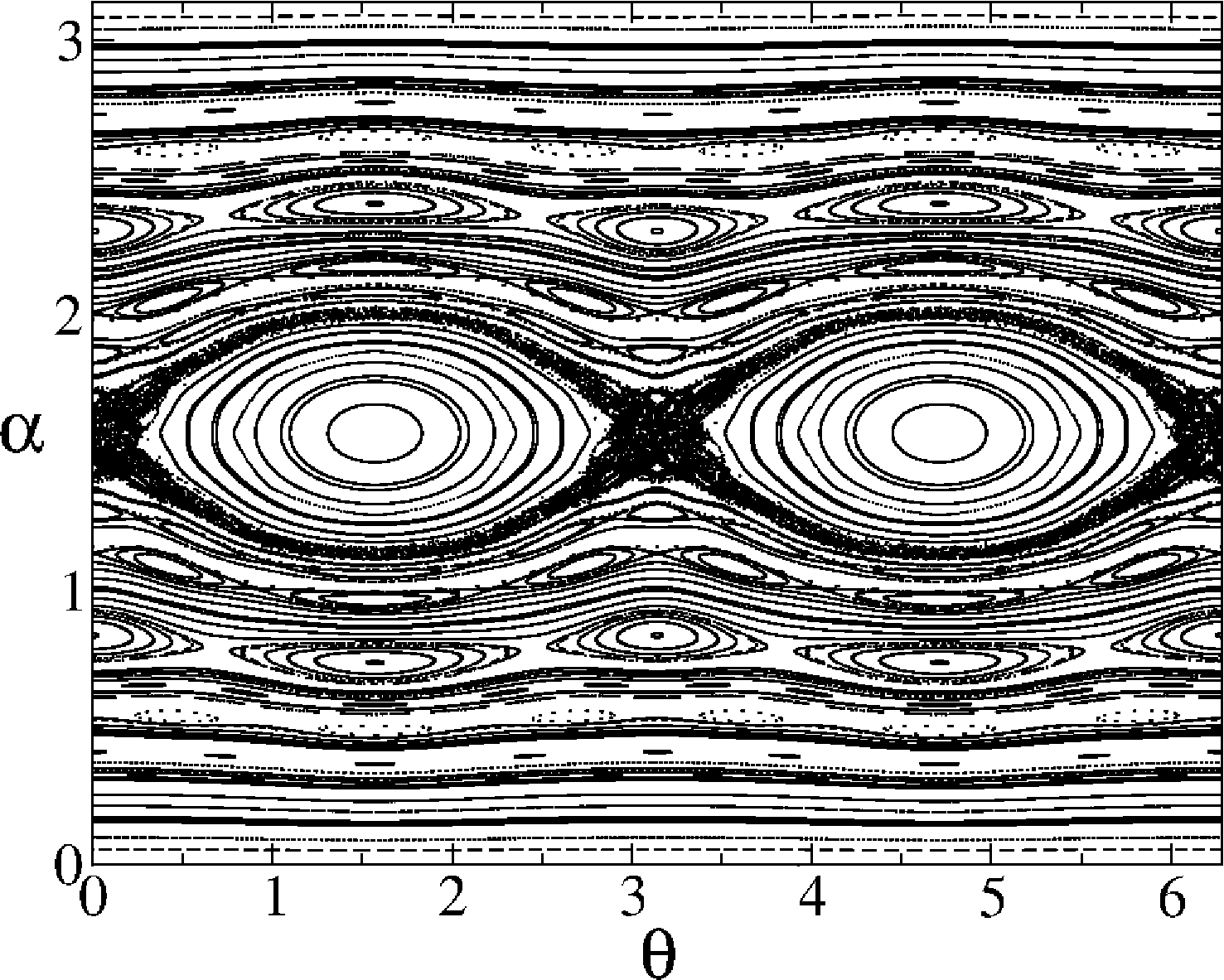}
(d)\includegraphics[width=0.47\linewidth]{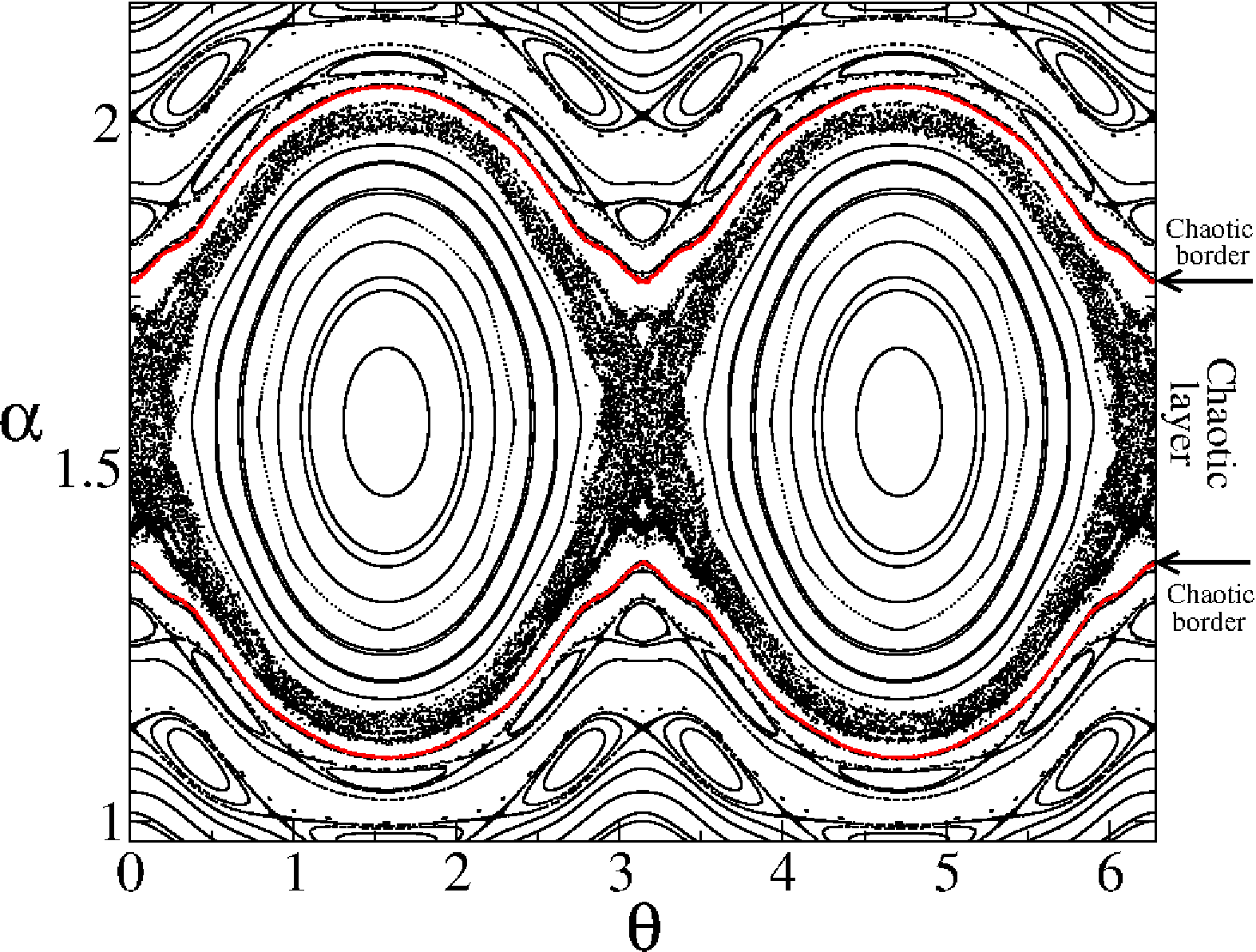}}
\caption{(Color online) (a) Sketch of the angles describing the dynamics of the billiard. The boundary was constructed using $R=1+\epsilon\cos(p\theta)$ with $p=2$ (assumed constant in this Letter) and $\epsilon=0.1$. In the figure, $\theta_0$ denotes the polar angle at the initial collision, $\vec{T}_0$ is the tangent vector at $\theta_0$, and $\alpha_0$ is the angle of the trajectory measured with respect to $\vec{T}_0$. (b) Phase space for $\epsilon=0$ showing a symmetric and ordered dynamics. (c) Phase space for $\epsilon=0.05$, displaying a mixed structure with a chaotic layer near $\alpha=\pi/2$, periodic islands and invariant spanning curves. The periodic domain confined by the chaotic layer corresponds to librational motion, while the invariant spanning curves above and below it correspond to rotational motion (whispering-gallery orbits \cite{wisp0,wispering}). (d) Zoom of (c) showing the chaotic domain bounded by the upper and lower chaotic borders (continuous red curves).}
\label{Fig1}
\end{figure}

The model we consider is the oval billiard \cite{R8}, whose boundary in polar coordinates is given by $R(\theta)=1+\epsilon\cos(p\theta)$, where $\theta$ is the polar angle, $\epsilon$ controls the deformation from the circle, and $p$ is an integer. The dynamics of a particle confined within the boundary is described by a discrete mapping for the variables $(\theta_n,\alpha_n)$, where $\theta_n$ denotes the angular position of the particle at the $n$th collision and $\alpha_n$ is the angle between the particle's trajectory and the tangent vector to the boundary at $\theta_n$, as illustrated in Fig.~\ref{Fig1}(a).

Since the boundary is written in polar coordinates, the position of the particle in rectangular coordinates at collision $n$ is
$X(\theta_n)=R(\theta_n)\cos(\theta_n)$ and 
$Y(\theta_n)=R(\theta_n)\sin(\theta_n)$. Given an initial condition $(\theta_n,\alpha_n)$, the angle between the tangent vector at the boundary point $(X(\theta_n),Y(\theta_n))$ and the horizontal axis is
$\phi_n=\arctan\!\left[Y'(\theta_n)/X'(\theta_n)\right]$. In the absence of potential gradients between collisions, the particle moves in a straight line with constant velocity. The trajectory between collisions satisfies the equation
\begin{eqnarray}
Y(\theta_{n+1}) - Y(\theta_n) = \tan(\alpha_n + \phi_n)\left[
X(\theta_{n+1}) - X(\theta_n) \right],
\label{B_eq2}
\end{eqnarray}
where $X(\theta_{n+1})$ and $Y(\theta_{n+1})$ are the rectangular coordinates of the next collision point obtained by solving numerically Eq. (\ref{B_eq2}). The angle between the trajectory and the tangent vector at $\theta_{n+1}$ is then
$\alpha_{n+1} = \phi_{n+1} - (\alpha_n + \phi_n)$.

The mapping describing the particle dynamics is written as
\begin{equation}
\left\{
\begin{array}{ll}
H(\theta_{n+1}) = R(\theta_{n+1})\sin(\theta_{n+1}) - Y(\theta_n) - \\
~~~~~~\tan(\alpha_n + \phi_n)\left[ R(\theta_{n+1})\cos(\theta_{n+1}) 
- X(\theta_n) \right], \\
\alpha_{n+1} = \phi_{n+1} - (\alpha_n + \phi_n),
\end{array}
\right.
\label{B_eq4}
\end{equation}
where $\theta_{n+1}$ is obtained numerically from $H(\theta_{n+1})=0$ and 
$\phi_{n+1}=\arctan[Y'(\theta_{n+1}) / X'(\theta_{n+1})]$. Figure~\ref{Fig1}(b) shows the phase space for $\epsilon=0$, corresponding to the circular billiard \cite{r1}. Besides energy conservation, the dynamics preserves the angular momentum, providing the two constants of motion required for integrability \cite{nivaldo}. The symmetry of the phase space, with constant reflection angle $\alpha$, is clearly observed. In contrast, Fig.~\ref{Fig1}(c) displays the phase space for $\epsilon=0.05$, where the mixed structure emerges: chaotic regions, stability islands, and quasiperiodic trajectories. The symmetry present in Fig.~\ref{Fig1}(b) is broken in Fig.~\ref{Fig1}(c), confirming that symmetry breaking accompanies the transition.

The chaotic regime develops within a well-defined layer in phase space, whose width depends on the strength of the control parameter $\epsilon$ and is bounded by two limiting curves, as shown in the zoom of Fig.~\ref{Fig1}(d). The width of this chaotic band increases monotonically with $\epsilon$, providing a geometric measure of the growth of chaos as regularity is progressively destroyed.

\begin{figure}[t]
\centerline{(a)\includegraphics[width=0.47\linewidth]{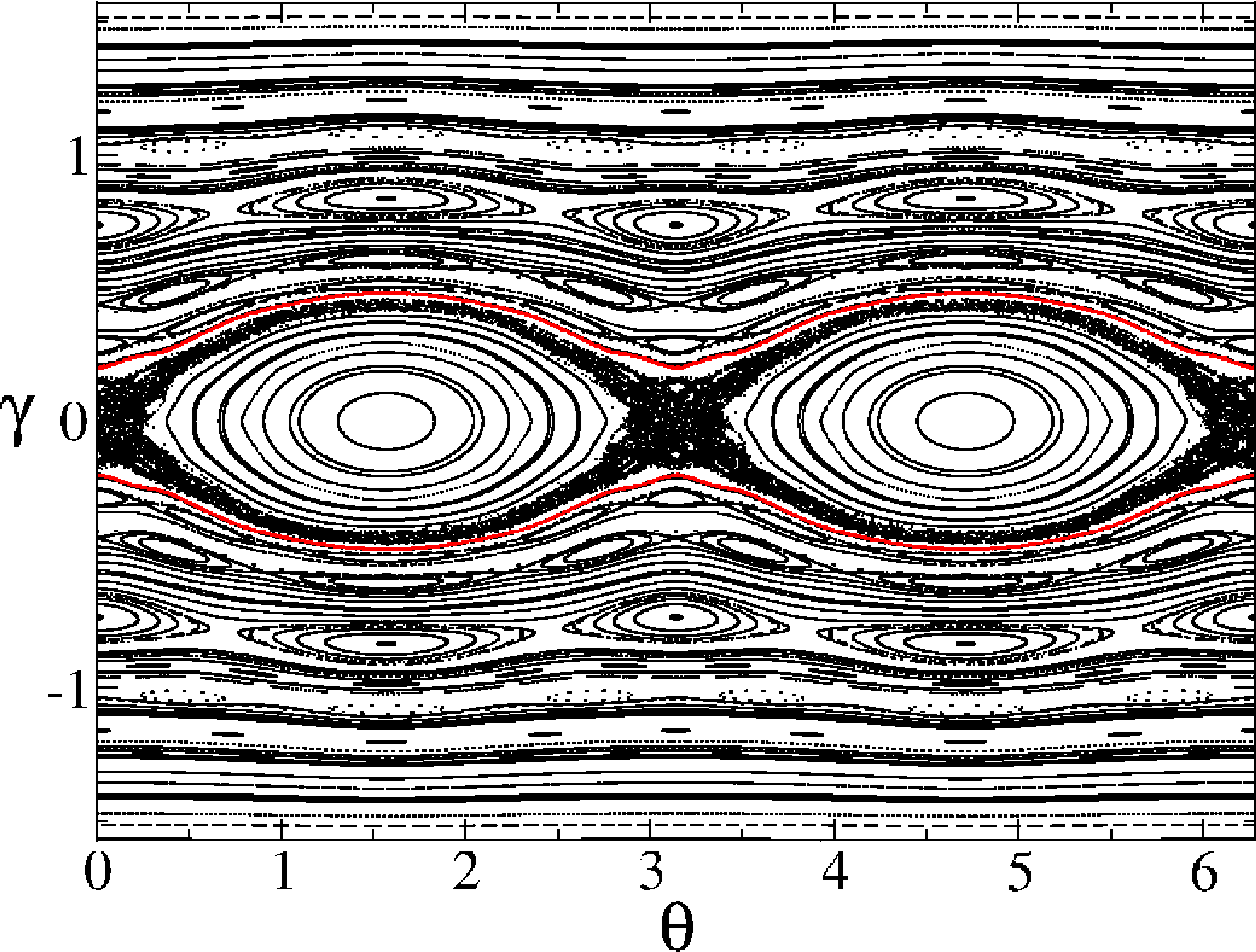}
(b)\includegraphics[width=0.47\linewidth]{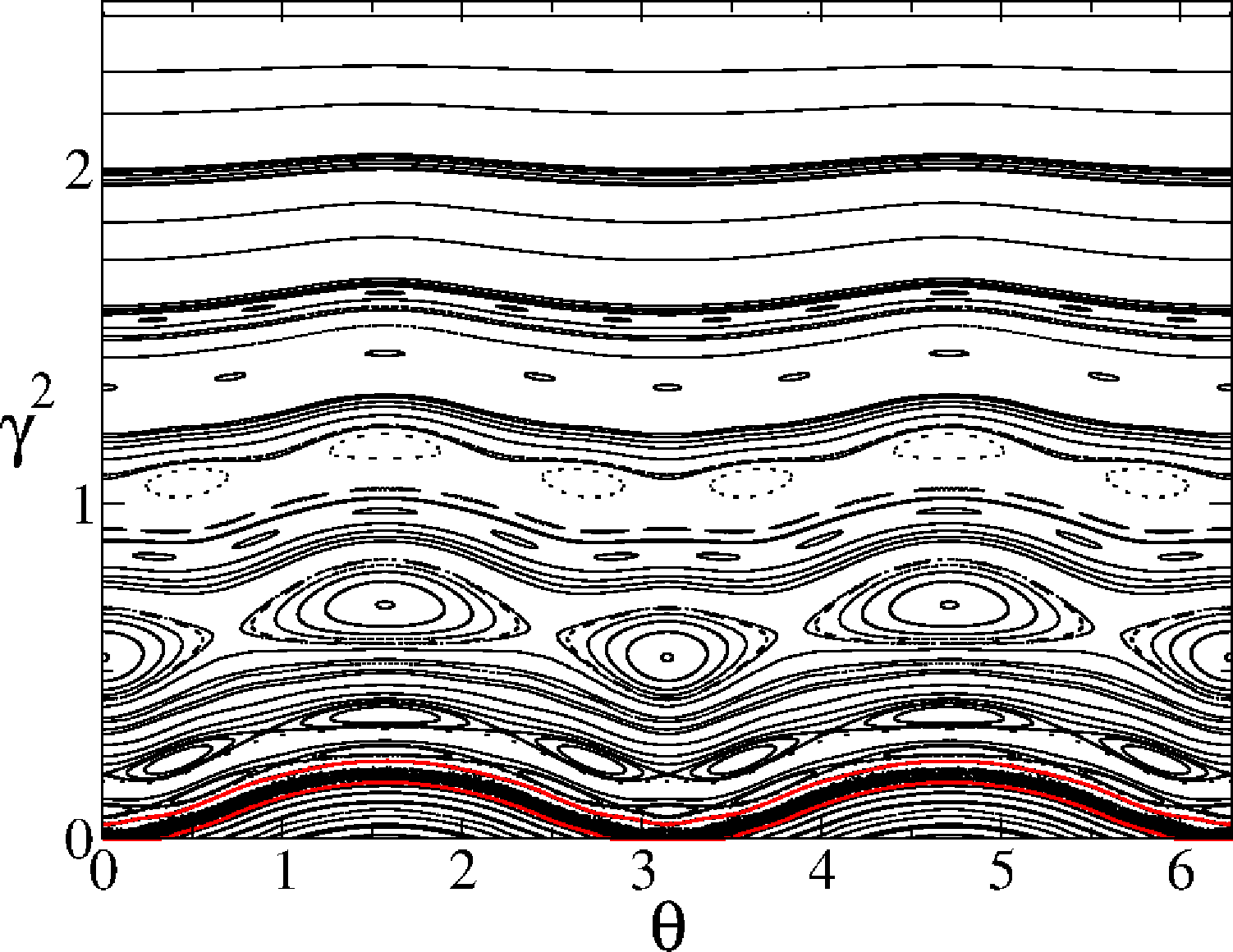}}\vspace{0.5cm}
\centerline{(c)\includegraphics[width=0.47\linewidth]{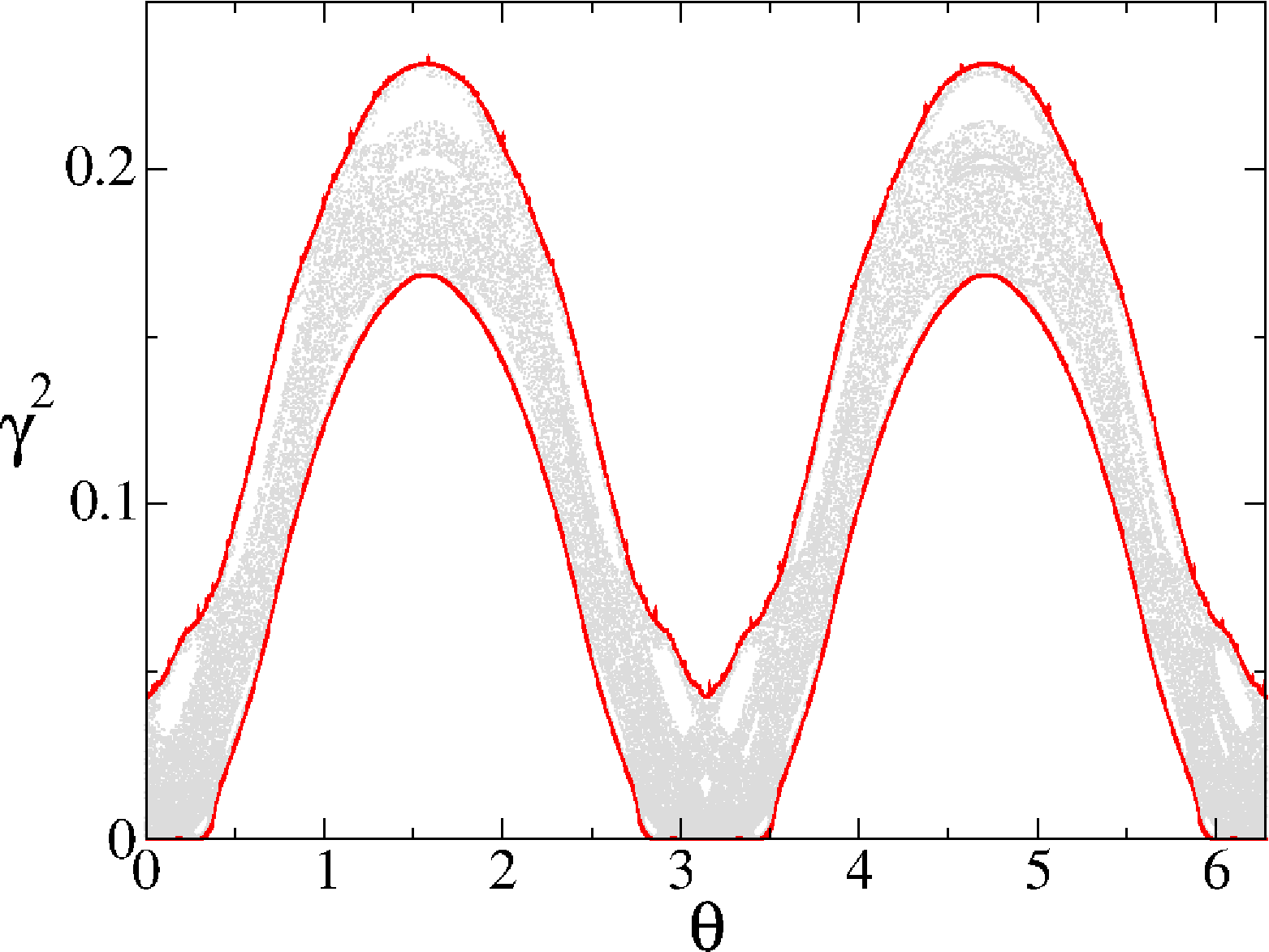}
(d)\includegraphics[width=0.5\linewidth]{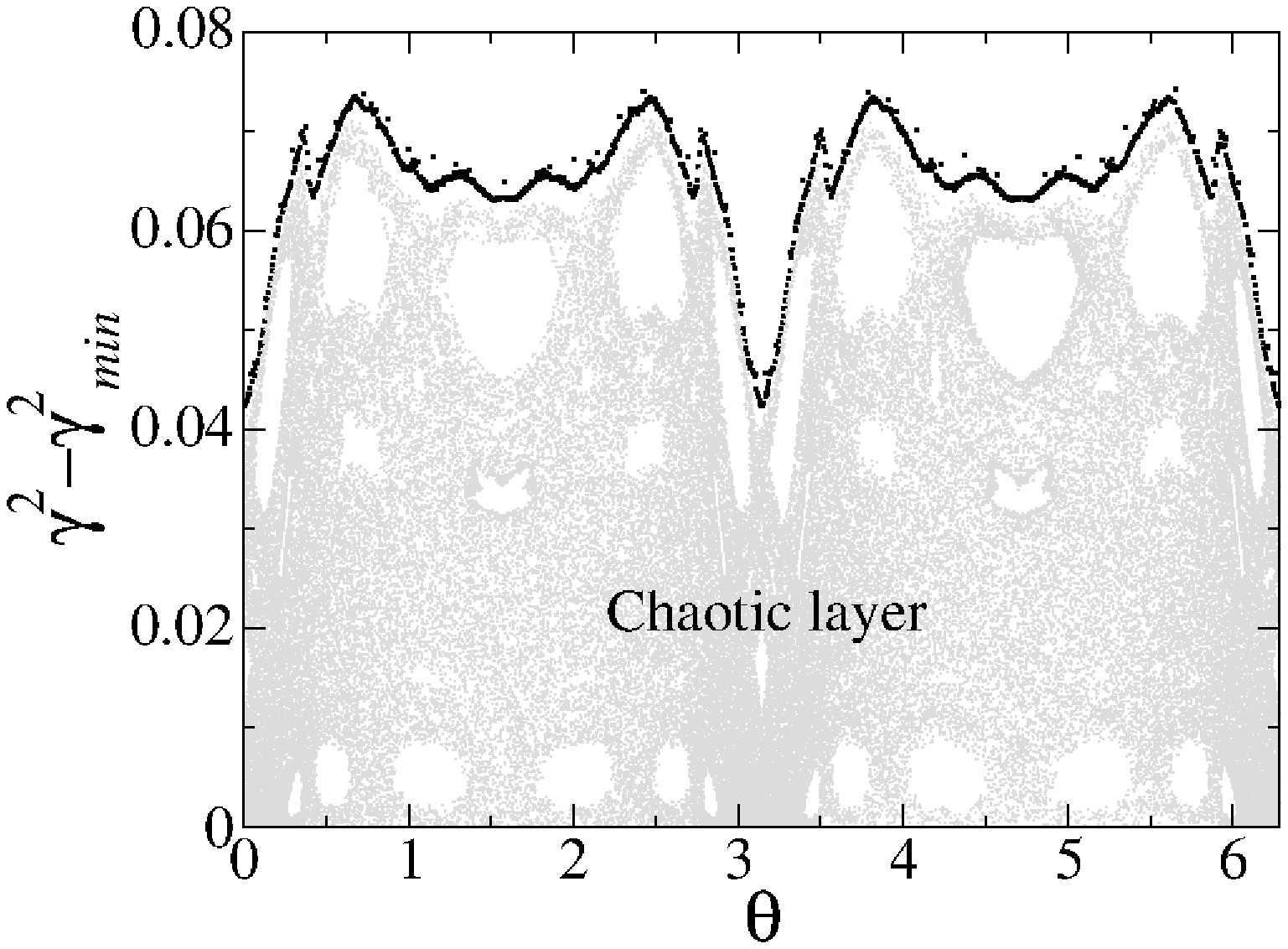}}
\caption{(Color online) (a) Phase space for $\epsilon=0.05$ under the transformation $\gamma=\alpha-\pi/2$. The phase space is now symmetric about $\gamma=0$, meaning that the positive part mirrors the negative part. (b) Plot of the square angle $\gamma^2$ vs.~$\theta$ for the same control parameter of (b). (c) The chaotic stripe in panel (b) bounded by two limiting curves. (d) Chaotic layer after the transformation $\gamma^2-\gamma_{min}^2$. The bullet points correspond to a numerical approximation of the chaotic border.}
\label{Fig2}
\end{figure}

To investigate the width of the chaotic layer, it is useful to introduce a more convenient set of variables. From Fig.~\ref{Fig1}(c), we observe that the phase space is symmetric with respect to $\alpha=\pi/2$, making it natural to shift this symmetry to the origin. We therefore define $\gamma=\alpha-\pi/2$, as illustrated in Fig.~\ref{Fig2}(a).

With this transformation, the phase space becomes symmetric about $\gamma=0$, so that the positive region is the mirror image of the negative one. Although this symmetry is conceptually useful, it implies $\bar{\gamma}=0$, which is not a convenient variable for measuring diffusion. Instead, we consider its square, $\gamma^2$, shown in Fig.~\ref{Fig2}(b). In these coordinates the entire phase space becomes positive, and the chaotic layer appears clearly as a band bounded by two curves, as shown in Fig.~\ref{Fig2}(c), defining a chaotic stripe. A further transformation, $\omega=\gamma^2-\gamma_{min}^2$, is particularly useful for analysing the width of the chaotic diffusion, as illustrated in Fig.~\ref{Fig2}(d).

As the control parameter $\epsilon$ increases, the width of the chaotic layer expands, forming a stripe along which chaotic motion develops. New periodic islands arise in the phase space as a consequence of the boundary deformation. The invariant spanning curves associated with grazing trajectories \cite{wisp0,wispering}, those with small angles nearly tangent to the boundary, become progressively less prominent. Eventually, these curves disappear entirely once the control parameter reaches the critical value $\epsilon_c=1/(1+p^2)$ \cite{diego}. Beyond this point, all invariant spanning curves are destroyed, allowing chaos to propagate through most of the accessible phase space and permeate the periodic islands that remain. Thus, as the chaotic layer grows, increasingly large portions of the phase space become chaotic, providing the ideal setting for identifying a dynamical phase transition. Our aim is to identify an observable along the chaotic diffusion that can serve as an order parameter \cite{r4}, distinguishing the ordered (integrable) phase from the disordered (chaotic) regime, and that approaches zero continuously at the transition. Simultaneously, the response of this order parameter \cite{r10} to variations in the control parameter must diverge in the limit where the nonlinearity $\epsilon$ vanishes.

\begin{figure}[t]
\centerline{(a)\includegraphics[width=0.8\linewidth]{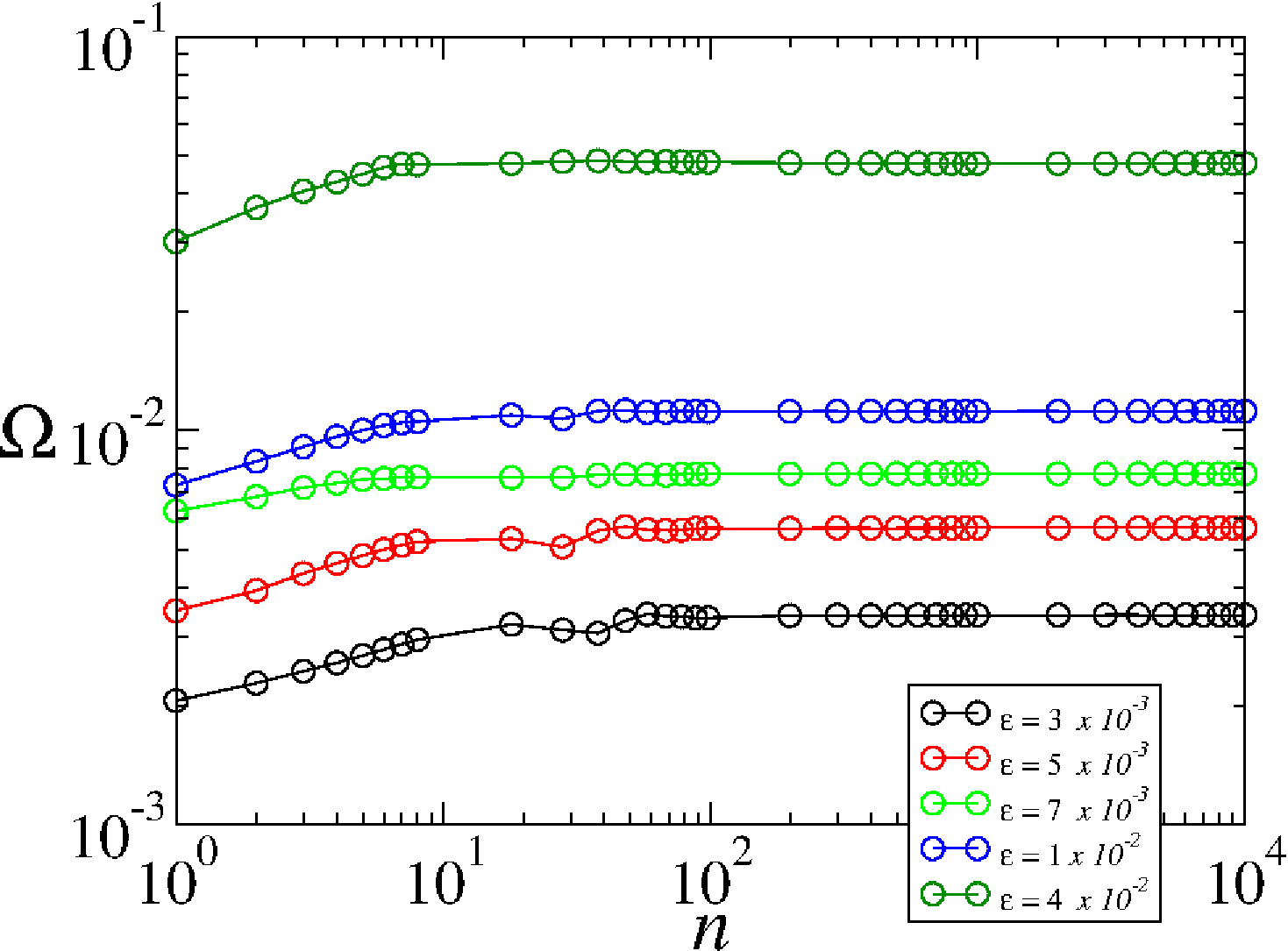}}
\centerline{(b)\includegraphics[width=0.75\linewidth]{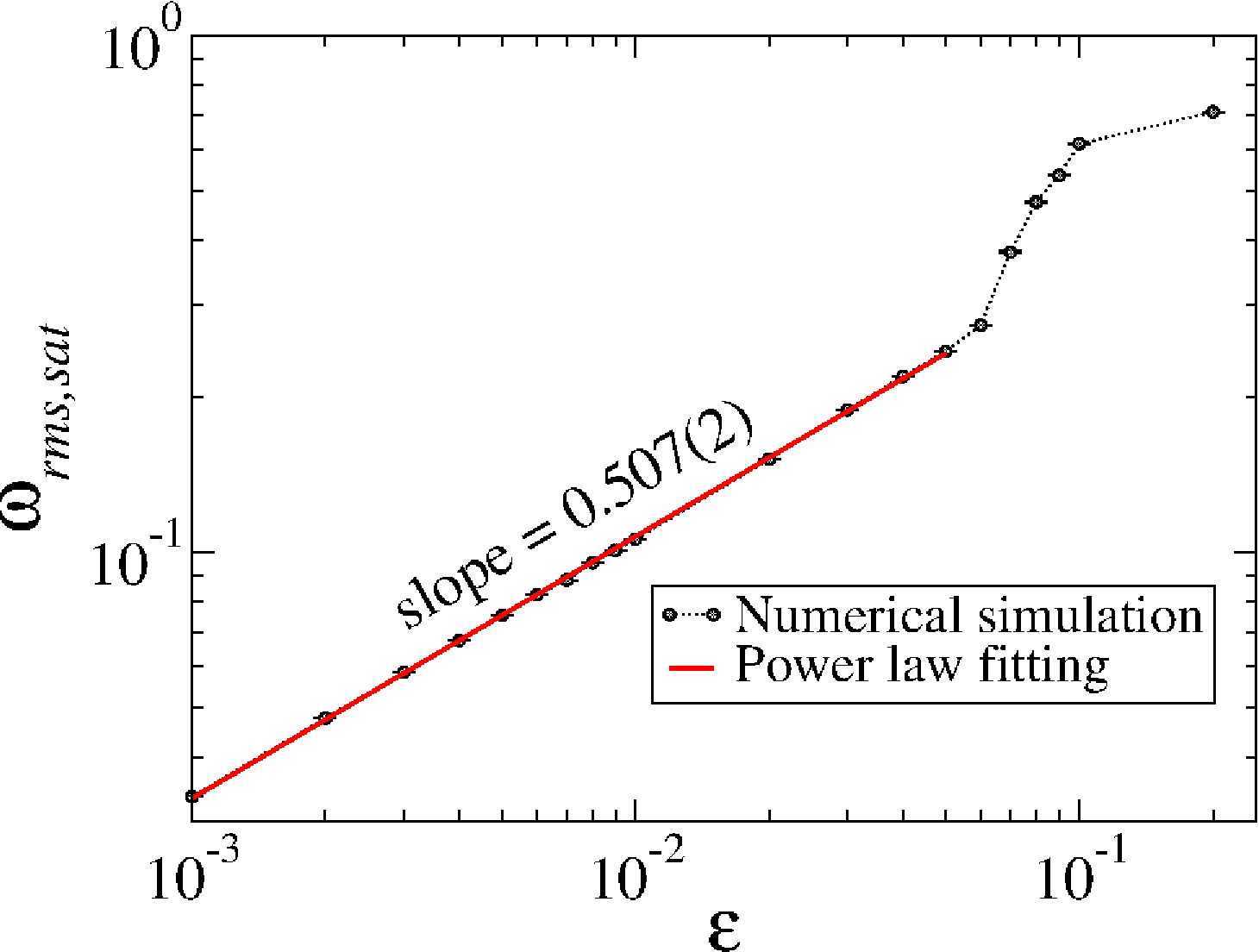}}
\caption{(Color online) (a) Plot of $\Omega~vs.~n$ for different control parameters, as labelled in the figure. (b) Plot of $\omega_{rms,sat}~vs.~\epsilon$. A power-law fit gives $\tilde{\alpha}=0.507(2)$. The error bars of the average measurements are also represented.}
\label{Fig3}
\end{figure}

We now investigate the behavior of the deviation around the chaotic stripe as a function of the control parameter $\epsilon$, an observable defined as 
\begin{equation}
\Omega(n,\epsilon)=\frac{1}{M}\sum_{i=1}^{M}\sqrt{\bar{\omega_i^2}(n,\epsilon)-\bar{\omega_i}^2(n,\epsilon)},
\end{equation}
where $\bar{\omega}(n,\epsilon)=\frac{1}{n}\sum_{k=1}^n\omega_k$. We consider an ensemble of $M=10^3$ different initial conditions with initial angles $\alpha$ chosen very close to the lower limit of the chaotic stripe and $\theta_0$ uniformly distributed in $\theta_0\in[0,2\pi]$, thus allowing the particles to experience maximum diffusion within the chaotic region. Figure \ref{Fig3}(a) shows a plot of $\Omega~vs.~n$ for different values of the control parameter $\epsilon$, as indicated in the figure. We observe that diffusion initially grows with the number of collisions and, after a crossover, eventually saturates, marking a steady state of the diffusive process.

Since $\Omega$ is defined in terms of the squared angle (it has units of squared angle $\alpha$), it is convenient to analyse the behavior of $\omega_{rms}=\sqrt{\Omega}$. For sufficiently large $n$, all curves for $\omega_{rms}$ approach a saturation regime. A direct measure of the saturation values for different $\epsilon$ is shown in Fig.~\ref{Fig3}(b). There is a clear power-law regime for small $\epsilon$, apart from the region near $\epsilon=0.07$ and larger, where the chaotic domain begins to grow rapidly in phase space and dominate the dynamics. This behavior occurs far from the vicinity of integrability, where the transition is characterised, and is not relevant for the critical analysis. A power-law fit for the initial curve, marking the points nearest to criticality, yields $\omega_{rms,{\rm sat}}\propto \epsilon^{\tilde{\alpha}}$ with $\tilde{\alpha}=0.507(2)$. Since we are analysing diffusive behavior, it is natural that the order parameter should quantify the extent of diffusion \cite{leonel3} along the chaotic stripe. The observable $\omega_{rms,{\rm sat}}$ satisfies all the necessary conditions to play this role. It is well defined within the chaotic stripe and, for small $\epsilon$ in the vicinity of the transition, it follows a power law in $\epsilon$ and as $\epsilon\rightarrow 0$, $\omega_{rms,{\rm sat}}\rightarrow 0$ in a continuous and monotonic way. On the other hand, the response of $\omega_{rms,{\rm sat}}$ to variations in the control parameter, which is precisely the susceptibility, is given by $\chi=\frac{d\omega_{rms,{\rm sat}}}{d\epsilon}\propto \frac{\tilde{\alpha}}{\epsilon^{\,1-\tilde{\alpha}}}$. Since $0<\tilde{\alpha}<1$, the susceptibility diverges as $\epsilon\rightarrow 0$. These results indicate that the transition from integrability to non-integrability in the oval-like billiard exhibits the characteristic features of a second-order phase transition.

Regarding the question of elementary excitations in the dynamics, when $\epsilon=0$, no diffusion is observed in phase space, which is entirely regular \cite{r1}. For small values of $\epsilon$, however, a chaotic layer appears and propagates along a stripe in phase space. This behavior is possible solely because the control parameter $\epsilon$ is nonzero. We therefore conclude that $\epsilon$ plays the role of an elementary step length that enables diffusive dynamics to unfold in phase space in the reflection angle $\alpha$.

Considering the possible topological defects that affect particle transport in phase space, from Fig.~\ref{Fig2}(d) we observe regions that remain empty, i.e., they are not visited by chaotic orbits and correspond to stability islands associated with periodic orbits. When trajectories wander close to such islands, they may become temporarily trapped before escaping, later returning to the region and becoming trapped again \cite{trap}. These episodes of stickiness, combined with the presence of stability islands, break ergodicity and modify the probability distribution of particles evolving under chaotic dynamics. However, because the chaotic layer has a finite size, bounded by the minimum and maximum extent of chaotic motion, the deviation around the average angle eventually saturates for long times. As discussed in Ref.~\cite{r15}, the presence of periodic islands is equivalent to the presence of topological defects that affect the diffusion of particles in phase space.

The diffusion along the chaotic stripe can also be described via the solution of the diffusion equation \cite{r11}. The probability of observing a particle with angle $\omega=\gamma^2-\gamma_{min}^2$ at a specific time $n$ is governed by ${{\partial P}\over{\partial n}}=D{{\partial^2P}\over{\partial\omega^2}}$ where $D$ is the diffusion coefficient obtained from
\begin{equation}
D=\lim_{n\to\infty}{{D_n}\over{n}},
\end{equation}
where
\begin{equation}
D_n=\sum_{i=1}^M<(\omega^i_n-\omega^i_0)^2>,
\end{equation}
where $M$ is the size of the ensemble of initial conditions. Since particles cannot cross the edges of the chaotic layer, the boundary conditions are 
\begin{equation}
\left.{{\partial P}\over{\partial\omega}}\right|_{\omega=(0,\omega_{fisc})}=0
\end{equation}
with the initial condition $P(\omega,0)=\delta(\omega-\omega_0)$. Here $\omega_{fisc}$ corresponds to the position of the upper curve bounding the chaotic diffusion. The solution \cite{r11} is given by
\begin{equation}
P(\omega,n)={{1}\over{\omega_{fisc}}}+{{2}\over{\omega_{fisc}}}
e^{-{Dm^2\pi^2n}\over{\omega_{fisc}^2}}\left[b_m\cos\left({{m\pi\omega}\over{\omega_{fisc}}}\right)\right],
\end{equation}
with $m=1,2,3,\ldots$ where $b_m=\cos\left({{m\pi\omega_0}\over{\omega_{fisc}}}\right)$. The time-dependent term vanishes as $n\to\infty$ and corresponds to the transient dynamics. The stationary state is given by $P(\omega,n\to\infty)={{1}\over{\omega_{fisc}}}$. Therefore the existence of a curve $\omega_{fisc}$ as an upper border for the diffusion allows the determination of any moment of the probability distribution. The saturation of $\omega_{sat}$ is a consequence of the existence of $\omega_{fisc}$.

As a short discussion, a phase transition in a family of area-preserving mappings \cite{leonel3} of the type
\begin{equation}
\left\{\begin{array}{ll}
I_{n+1}=I_n+\epsilon \sin(\theta_n),\\
\theta_{n+1}=[\theta_n+{{1}\over{|I_{n+1}|^{\tilde{\gamma}}}}]~~{\rm mod~(2\pi)},\\
\end{array}
\right.
\label{eq1}
\end{equation}
with $\tilde{\gamma}>0$, shows that the diffusive behavior saturates due to the existence of invariant spanning curves in phase space. Their position is estimated as
\begin{equation}
I_{fisc}\cong \left[{{\tilde{\gamma}\epsilon}\over{0.9716\ldots}}\right]^{{1}/{(1+\tilde{\gamma})}}.
\label{fisc}
\end{equation}
The curves given by Eq. (\ref{fisc}) play exactly the same role as $\omega_{fisc}$ in the present model. The case $\tilde{\gamma}=1$ recovers both the Fermi-Ulam model \cite{leonel1} and the periodically corrugated waveguide \cite{leonel2}. The critical exponent governing the diffusive behavior is $\tilde{\alpha}=1/(1+\tilde{\gamma})$. For $\tilde{\gamma}=1$, the critical exponent is $\tilde{\alpha}=1/2$, remarkably the same value observed in the transition reported in the present work. Therefore, we conclude that, even though the systems considered are quite distinct, whether a Fermi-Ulam model \cite{leonel1}, a periodically corrugated waveguide \cite{leonel2}, or a family of area-preserving mappings \cite{leonel3} such as that defined in Eq. (\ref{eq1}), the phase transition observed here belongs to the same universality class as that of the Fermi-Ulam model.

The procedure used here consists of:
(1) identifying a symmetry breaking; 
(2) defining an order parameter and its susceptibility and analysing their behavior at the transition; 
(3) searching for topological defects; and 
(4) identifying an elementary excitation that drives diffusion. 
This procedure can be applied broadly to investigate different types of phase transitions in distinct areas of science, making the approach of wide and general interest.

E.D.L. acknowledges support from Brazilian agencies CNPq (No. 301318/2019-0, 304398/2023-3) and FAPESP (No. 2019/14038-6 and No. 2021/09519-5).

\end{document}